\documentstyle[twoside,fleqn,espcrc2]{article}

\newcommand{\AmS}{{\protect\the\textfont2
  A\kern-.1667em\lower.5ex\hbox{M}\kern-.125emS}}

\title{Do large abelian monopole loops survive the continuum limit?}

\author{M. Grady\address{Department of Physics, 
        SUNY Fredonia, Fredonia NY 14063 USA}}
       
\begin{document}

\begin{abstract}
An analysis of the monopole loop length distribution is performed in
Wilson-action SU(2) lattice gauge theory. A pure power law in the inverse 
length is found, 
at least for loops of length, $l$, less than the linear lattice size $N$.
This power shows a definite $\beta$ dependence, passing 5 around $\beta=2.9$,
and appears to have very little finite lattice size dependence. It
is shown that when this power exceeds 5, no loops any finite fraction
of the lattice size will survive the infinite lattice limit. This is true
for any reasonable size distribution for loops larger than N.  The apparent
lack of finite size dependence in this quantity would seem to indicate
that abelian monopole loops large enough to cause confinement do not
survive the continuum limit. Indeed they are absent for all $\beta > 2.9$.
\end{abstract}

\maketitle

\section{INTRODUCTION}
Much evidence has been presented that abelian monopoles extracted from
SU(2) lattice gauge theory in the maximal abelian gauge are related
to the confinement mechanism\cite{mag}. 
The monopoles appear to carry the entire
SU(2) string tension. Long loops of monopole current at least as large
as the lattice appear to be necessary for confinement. 
Evidence has also been
presented that these monopoles survive 
the continuum limit\cite{poli}. 
It is important to ask whether these surviving 
monopoles exist in large loops capable of causing confinement, or  
only in small loops which are of zero physical
size in the continuum.

\section{MONOPOLE LOOP DISTRIBUTION}

SU(2) Wilson-action lattices are transformed to maximal abelian gauge and
abelian monopoles extracted using the DeGrand-Toussaint 
procedure\cite{dg}. The number of loops 
of each length are tabulated. 
It is found that small loops are more common than 
large loops. At larger
$\beta$ this effect is strongly enhanced. Define $p(l)$ to 
be the probability,
normalized per lattice site, of a loop of size $l$ 
occurring on a lattice. The quantity
$N^4 p(l)$ gives the average number of loops of size $l$ occurring 
on an $N^4$ lattice. 
In Fig.~1 log$_{10} P(l)$ is plotted vs. log$_{10}(l)$.
\begin{figure}[ht]
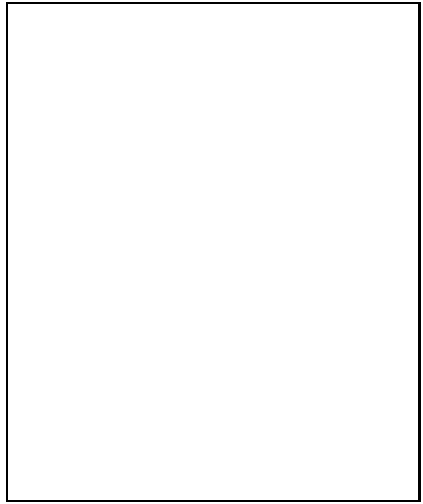

\framebox[55mm]{\rule[-32mm]{0mm}{64mm}}
\caption{Log-log plots of loop probability vs. loop size. Lines 
(from upper to lower) are fits
to $12^4$ data for $\beta=$ 2.6,2.7,2.8,2.9, and 3.0}
\label{fig:fig1}
\end{figure}
It is seen that in all cases a power law is followed for smaller loops,
up to a point somewhat beyond $l=N$. Beyond this there is a bulge of excess
probability, followed by a sharp drop. This bulge is easily understood as
a finite size effect due to the periodic boundary condition. Would-be large
loops can reconnect through the boundary; thus one obtains
an excess of mid-size loops ($l>N$) but a deficit of very large loops.
This bulge is not so apparent for larger $\beta$. At any rate, 
the loop distribution certainly appears to follow a pure
power law for $l \leq N$. Similar results were obtained in \cite{teper}.
This power appears to have hardly any finite lattice size dependence,
with the $12^4$ data lying very close to the $20^4$. The main difference
in the data is a slight shift in the intercept.
Indeed, since the 
power is already established by loops of length  6 and 8, it is hard
to imagine how it could differ on very large lattices 
from what is seen here for the $20^4$ lattice, especially considering
how little difference there is between the $12^4$ and $20^4$ data.
It seems unlikely that these small loops would be 
sensitive to the overall lattice size. Thus the power 
will likely be the same on very large lattices
as observed here for the $20^4$ lattice.
It should be noted that
minimal
loops of size 4 do not fall on the power law line and are excluded from
fits, and, with good statistics one can also see 
that the size 6 loops fall very slightly below the power
trend.
Taking the power law to be $p(l) \propto l^{-q}$, the power $q$ obtained
from linear fits is plotted
vs. $\beta$ in Fig.~2. A strong $\beta$ dependence is apparent. 
This is in contradistinction to Ref.~\cite{teper}, where it was
assumed there is no $\beta$ dependence in this quantity.
\begin{figure}[t]
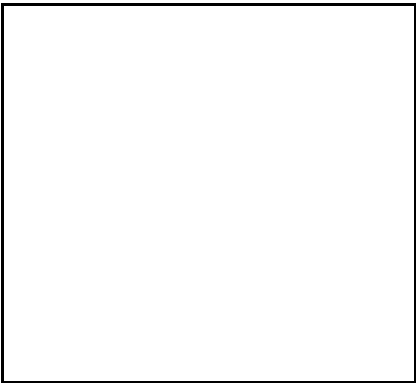

\framebox[55mm]{\rule[-24mm]{0mm}{48mm}}
\caption{The power, $q$ vs. $\beta$.}
\label{fig:fig2}
\end{figure}
Above $\beta=2.85$, $q>5$, the significance of which is 
explained below. 

Consider the probability of finding a loop of any size between 
$N/b$ and $N$ on an $N^4$ lattice, where $b$ is fixed.
For large $N$ and $b<<N$, this is given by $N^4	I(N,b)$ where 
\begin{equation}
I(N,b)=\int^{N}_{N/b} p(l) dl .
\end{equation}
Here a sum over lengths 
has been replaced by an integral since $N$ and $N/b$ are large.
Taking $p(l) \propto l^{-q}$, $N^4 I(N,b) \propto N^{5-q}$.  Thus for
$q>5$, the probability of any loop in the size range $N/b$ to $N$ existing
on an $N^4$ lattice for any fixed $b$ vanishes as $N \rightarrow \infty$.
If there are no loops in this range, it seems unlikely
that larger loops could occur. Clearly, if the power law continues
they do not. However, even if the loop distribution falls more slowly
for $l>N$, such loops will not occur in the infinite lattice so long as the
loop distribution continues to fall by a power $\geq 1$. It would be
nearly impossible for it to fall more slowly than this. For instance,
assuming the distribution follows a different power law for $l>N$ (for
which there is no evidence) it would look like 
$p(l) \propto N^{-(q-q')} l^{-q'}$
for $l>N$, where $q$ is the power for $l<N$. The probability for
having a loop of size $l>N$ on a lattice is $N^4 \int^{8N^{4}}_{N} p(l) dl$.
It is easy to see that for $q>5$ and $q' \geq 1$ this quantity vanishes
for $N \rightarrow \infty$. Thus, somewhat paradoxically,
the probability of large
loops on large lattices is controlled by the probability distribution
for $l<N$, where its behavior appears to be a pure power law,
the power for which is established in turn 
by the behavior of very small loops.

To sum, what has been shown is that if the power, $q$, of loop probability 
falloff with
loop length is larger than 5 in the region $l<N$, and larger than unity
for $l>N$,
then no loops any finite fraction of the 
lattice size exist in the infinite lattice limit. 
Previously it was seen from Fig.~2 that  
$q>5$ for all $\beta \geq 2.9$.
The apparent lack of dependence on lattice size for $q$ 
means that this result should continue to hold for a very large and even
infinite lattice. One is therefore led to the conclusion that
monopole loops large enough to cause confinement do not survive the
continuum ($\beta \rightarrow \infty$) limit.  Therefore, if SU(2)
lattice gauge theory confines in this limit, then confinement is
not due to abelian monopole loops in the continuum.  Another possibility
is to accept that abelian monopole loops do cause confinement, in 
which case one is led to the conclusion that the continuum limit
of SU(2) lattice gauge theory is not confined. Needless to say, either
conclusion is highly unconventional.
  
The monopole loops that do possibly survive the continuum limit are all
of zero physical size, so are unlikely to affect 
the continuum theory in any way.
This can be seen by taking a large but finite universe. 
Then the
continuum limit can be taken by taking the lattice spacing 
$a\rightarrow 0$
and $N\rightarrow \infty$ together, keeping $Na$ fixed 
at the universe
size. Any physical size, such as that
of a hadron,
is a finite fraction of the universe size, $Na/b$, 
where b is finite. 
It was found above that no loops any finite 
fraction of the
lattice size survive the continuum limit. The largest loop 
on a typical lattice
must scale slower than $N$. 
Since $a$ scales as $1/N$,
such objects shrink to zero physical size in the continuum limit.

This suggests another way to analyze the data. 
One can measure the average largest loop (ALL) as a 
function of $N$ and $\beta$.
It is found that for $\beta < 2.9$, the ALL grows 
faster than $N$ as $N$
increases. Thus if wrapping loops are not present 
on a small lattice at
such $\beta$ they will eventually become 
common for large enough $N$. This
is one way to understand why small lattices are deconfined and large
lattices confine at the same $\beta$, 
and also why the critical beta of the deconfinement
transition depends so much on $N$ (see also \cite{wloop}).  
However, for $\beta>2.9$ it is 
found that the ALL scales {\em slower} 
than $N$ in going from a $12^4$ to a $20^4$
lattice.  Specifically, at $\beta=3.0$, ALL$/N=0.463 \pm 0.004$ 
on the $12^4$
and ALL$/N = 0.418 \pm 0.005$ on the $20^4$ lattice  
(errors are from binned fluctuations).  
Such behavior, of course,
follows from the fact that $q>5$ 
in the loop
distribution. If this trend continues, then for these values of
$\beta$ the probability of a wrapping loop will {\em decrease} as $N$
increases, and the confinement transition\footnote{
It is, of course, not a true phase transition on a finite lattice.} 
will never 
occur, no
matter how large N gets.  The picture that emerges from 
this analysis is
that the deconfinement transition does not, in fact continue to 
$\beta \rightarrow \infty$ as $N \rightarrow \infty$ 
but rather gets
stuck around $\beta = 2.9$, and becomes a bulk transition 
on the infinite
symmetric lattice, leaving 
the $\beta \rightarrow \infty$ continuum limit deconfined.

\section{SO(3)-Z$_2$ Monopoles}

The possibility that the zero-temperature 
continuum limit is deconfined
has been suggested before\cite{nccl}. 
It is further strengthened by
simulations which suppress a 
certain lattice artifact, 
the SO(3)-Z$_2$ monopole\cite{so3}.
This monopole can be defined as a nontrivial realization of 
the lattice Bianchi identity\cite{lbi}, 
and carries both an SO(3) and Z$_2$ charge.
It is an artefact because it is a local 
object of the scale
of a single lattice spacing which requires large-angle plaquettes
for support.  Such objects cannot exist in the continuum,
so an action which suppresses them should fall in the same
universality class as the Wilson action. However, simulations with this
action are deconfined for all $\beta$, despite 
the fact that this includes a region of rather strong renormalized
coupling (from the average plaquette), indicating that 
hadronic scales have
probably been reached. 
Analysis of the interquark potential for this action is underway, 
which should further elucidate this matter.

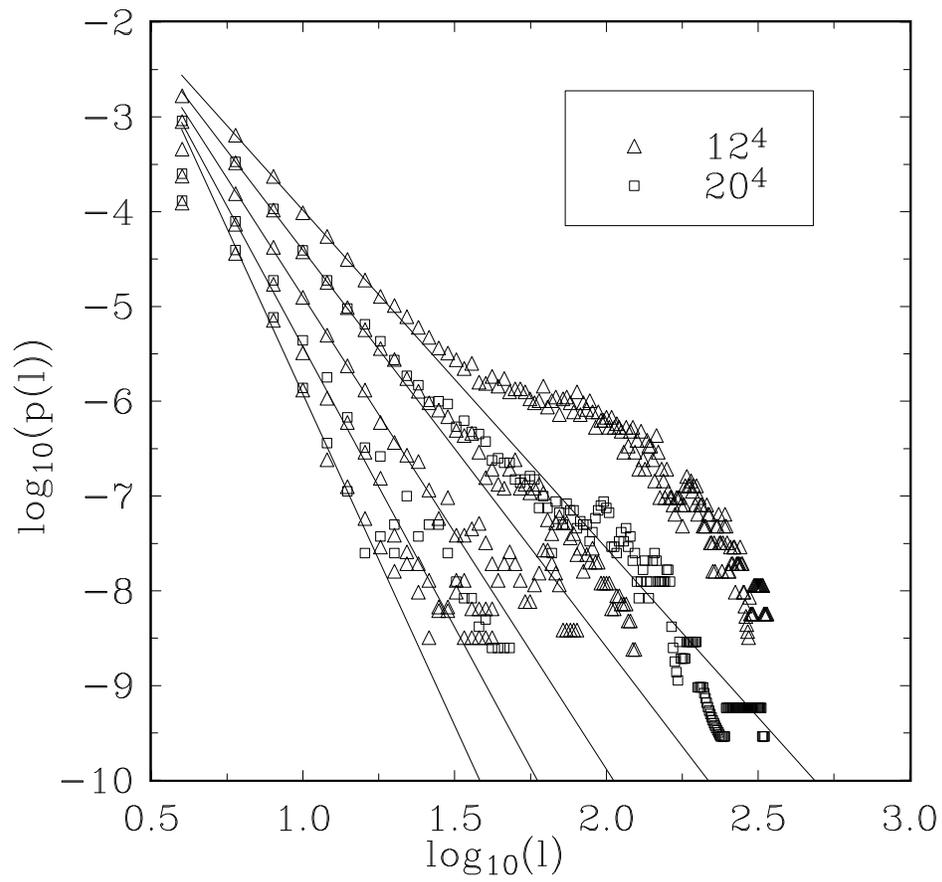
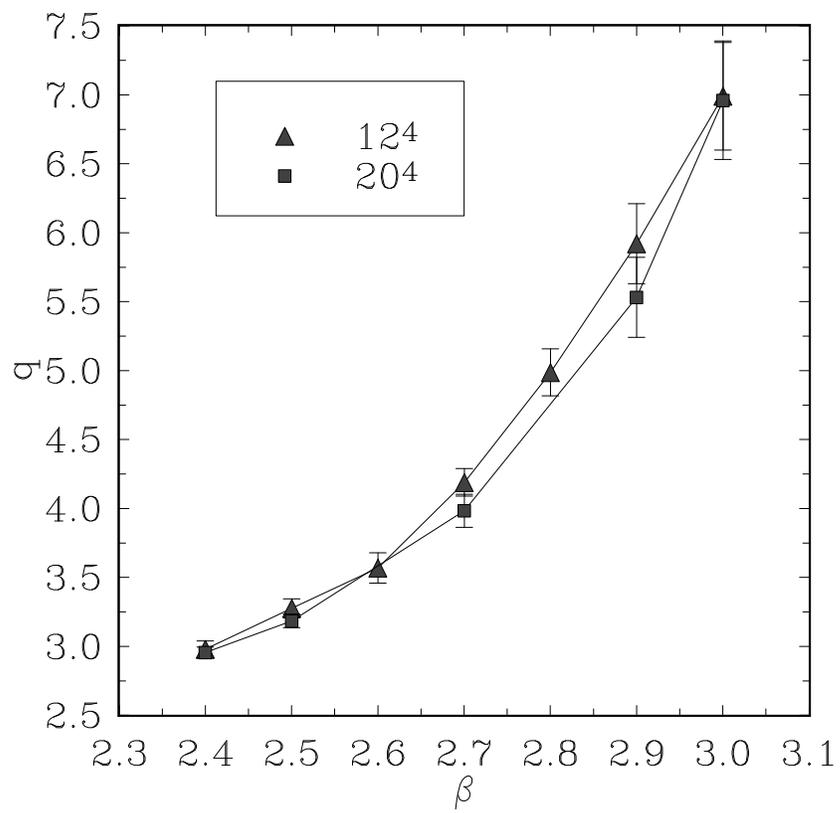
\end{document}